\documentclass[12pt,preprint]{aastex}

\newcommand{\hd}{HD~189733}
\newcommand{\xmm}{{\em XMM-Newton}}
\newcommand{\chandra}{{\em Chandra}}

\slugcomment{}

\shorttitle{The corona of \hd.\ }
\shortauthors{Pillitteri et al.}

\begin{document}


\title{The corona of \hd\ and its X-ray activity.}


\author{I. Pillitteri\altaffilmark{1,2}} 
\author{S. J. Wolk\altaffilmark{1}}
\author{J. Lopez-Santiago\altaffilmark{3}}
\author{H. M. G\"unther\altaffilmark{1}}
\author{S. Sciortino\altaffilmark{2}}
\author{O. Cohen\altaffilmark{1}}
\author{V. Kashyap\altaffilmark{1}}
\author{J. Drake\altaffilmark{1}}
\affil{SAO - 60 Garden st, Cambridge, MA 02138 - USA}
\affil{INAF-Osservatorio Astronomico di Palermo, Piazza del Parlamento 1, 90134 Palermo, Italy}
\affil{AEGORA, Facultad de CC. Matem\'aticas, Universidad Complutense de Madrid, 
Plaza de Ciencias 3, 28040, Madrid, Spain}


%

\begin{abstract}
Testing whether close-in massive exoplanets (hot Jupiters) can enhance the stellar activity in their host
primary is crucial for the models of stellar and planetary evolution. 
Among systems with hot Jupiters, \hd\ is one of the best studied because of 
its proximity, {strong activity and the presence of a transiting planet, that allows
transmission spectroscopy, a measure of the planetary radius and its density.} 
Here we report on the X-ray activity of the primary star, \hd~A, using a new 
\xmm\ observation and a comparison with the previous X-ray observations.
The spectrum in the quiescent intervals is described by two temperatures at 0.2 keV and 0.7 keV, while
during the flares a third component at 0.9 keV is detected. With the analysis of the summed RGS spectra,
we obtain estimates of the electron density in the range $n_e = 1.6 - 13 \times 10^{10}$ cm$^{-3}$ and
thus the corona of \hd~A appears denser than the solar one. 
{For the third time, we observe a large flare that occurred just after the eclipse of the planet.
Together with the flares observed in 2009 and 2011, the events are restricted to a small planetary 
phase range of $\phi = 0.55-0.65$. 
Although we do not find conclusive evidence of a significant excess of flares after 
the secondary transits, we suggest that the planet might trigger such flares when 
it passes close to locally high magnetic field of the underlying star at particular 
combinations of stellar rotational phases and orbital planetary phases.
For the most recent flares, a wavelet analysis of the light curve 
suggests a loop of length of four stellar radii at the location 
of the bright flare, and a local magnetic field of order of 40-100 G, in agreement with 
the global field measured in other studies. 
The loop size suggests an interaction of magnetic nature between planet and star, 
separated by only $\sim8 R_*$.
The X-ray variability of \hd~A is larger than the variability of field stars and young Pleiades 
of similar spectral type and X-ray luminosity. 
We also detect the stellar companion (\hd~B, $\sim12\arcsec$ from the primary star) 
in this \xmm\ observation. Its very low X-ray luminosity 
($L_X = 3.4\times 10^{26}$ erg s$^{-1}$) confirms the old age of this star and of the binary system. 
The high activity of the primary star is best explained by a transfer of angular momentum from the 
planet to the star.}
\end{abstract}


\keywords{stars: activity -- stars: individual (HD189733) -- stars: coronae -- stars: magnetic fields 
-- stars: planetary systems}

\section{Introduction}
The significant fraction of massive exoplanets that orbit at few stellar radii (hot Jupiters)
of the host primary is a challenge for the models of evolution of such systems. 
Evidence of star-planet interaction (SPI) is, however, still a matter of debate.
To first order, hot Jupiters should affect their host stars through both tidal and 
magneto-hydrodynamical effects 
(cf. \citealp{Cuntz2000}; \citealp{Ip2004}). {Both effects should scale with 
 the separation ($d$) between the two bodies as $d^{-3}$ \citep{Saar2004}.} 
The interaction between the respective magnetic fields of the hot Jupiter and the star may be a source 
of enhanced activity that could manifest in X-rays. Transfer of angular momentum from the planet to the star
during the inward migration and circularization of the orbit might also affect the stellar dynamo efficiency
and thus the intensity of the coronal emission in X-rays.

{Evidence of chromospheric activity induced by hot Jupiters in individual cases has been reported
first by \citet{Shkolnik03}, with other cases investigated by \citet{Shkolnik05}, \citet{Catala2007},
\citet{Fares2010}, \citet{Fares2012}, \citet{Lanza09}, \citet{Lanza2010}, \citet{Lanza2011},
\citet{Gurdemir2012} and \citet{Shkolnik2013}.} 
On large sample, enhanced chromospheric activity in stars with hot Jupiters has been reported
by \citet{Krejcova2012}, based on the analysis of Ca H\&K lines. 
\citet{Kashyap08} showed that stars with hot Jupiters are statistically brighter in X-rays than 
stars without hot Jupiters. On average \citet{Kashyap08} observed an excess of X-ray emission by a factor of 4
in the hot Jupiter sample. A similar result has been reported by \citet{Scharf2010}, who shows
a positive correlation between the stellar X-ray luminosity and the mass of their hot Jupiters.
However, analysis by \citet{Poppenhager2010} and \citet{Poppenhager2011} warn against the biases
that could affect the results of \citet{Kashyap08} and \citet{Scharf2010}, 
and suggest that SPI can take place only in specific systems under favorable conditions.

\hd\ is composed of a K1.5V  star at only 19.3 pc from Sun, 
and an M4 companion at 3200 AU from the primary, orbiting on a plane perpendicular 
to the line of sight. The primary hosts a hot Jupiter class planet (HD~189733 b) at a distance of 
only 0.031 AU with an orbital period of $\sim 2.22$d \citep{Bouchy05}.
The proximity of \hd\ and its transiting planet allow us detailed observations from IR to X-ray bands,
making this one of the best characterized systems with a hot Jupiter. 
In X-rays \hd\ has been observed with \xmm, \chandra\ and {\em Swift}.
In \citet{Pillitteri2010,Pillitteri2011} (hereafter Paper I and II, respectively) 
we investigated the X-ray emission of \hd\ and discovered signs of tidal/magnetic SPI 
after the planetary eclipse.
In this paper we present the results of the third \xmm\ observation taken at the secondary
transit of the hot Jupiter in the \hd\ system.  

The paper is organized as follows: Sect. 2 presents the target and  summarizes our previous results,
Sect. 3 describes the \xmm\ observation and the data analysis, Sect, 4 reports our new results,  Sect. 5 
discuss them, in Sect. 6 we outline our conclusions.


\section{\hd: previous X-ray observations}
With \xmm in 2009 and 2011 we observed the eclipse of the planet by the parent star ($\phi\sim0.45-.6$, 
{being $\phi = 0$ the transit of the planet in front of the star}) with
the goal of studying star planet interactions (SPI).  
In both observations a bright flare occurred after the end of the eclipse of 
the planet, at essentially the same phase. 


{The beginning of the flares observed in HD~189733 in 2009 and 2011 are at phases $\phi \sim 0.54$ 
and $\phi \sim 0.52$, respectively, 
which coincides with the moment when a region $\sim75^{\rm o}$ leading the sub-planetary 
point emerges on the limb of the star (i.e. $\phi \sim 0.73-0.75$ respectively, see also Fig. \ref{phases}). 
\citet{Shkolnik08} find enhanced chromospheric activity at similar phases (0.7-0.8) opening to
a scenario in which a connecting magnetic tube may exist between planet and star and generates
enhanced chromospheric activity as modeled by \citet{Lanza09}.}
Overall, the flares are associated with a modest change of the mean 
plasma temperature from  $\sim0.5$ keV to $\le 1.0$ keV, and a flux change of about a factor two. 
We point out that the two bright flares observed after the secondary transit of the planet 
are the brightest flares observed in \hd\ in all \xmm\ and \chandra\ observations.  
They pass the test of \citet{Wolk2005} for flare detection, 
while other variability detected at the primary transits does not pass the definition of flares.
 
The non-detection of the M companion \hd B in X-rays indicates very low activity 
($L_X\le10^{27}$ erg/s, Paper I). 
This conclusion is supported by spectroscopy (\citealp{Guinan2013} and  \citealp{Santapaga2011}). 
\citet{Poppenhager2013} detected \hd B at a level of $\log L_X \sim 26.4$, 
and confirmed its low X-ray activity. 
In Paper I, we interpreted the difference of activity between \hd A and \hd B as the result of strong 
tidal interaction between the planet \hd b and the parent star that induced a comparatively 
fast rotation at a late age of the primary, causing an enhancement of its overall activity.
This conclusion agrees with other evidences of induced extra rotation in host stars 
caused by inward migration and tidal circularization of massive planet orbits. 
In such systems gyrochronogical method would assign a younger age to these stars \citep{Pont2009}.
 
\begin{figure}
\begin{center}
\includegraphics[width=0.9\columnwidth]{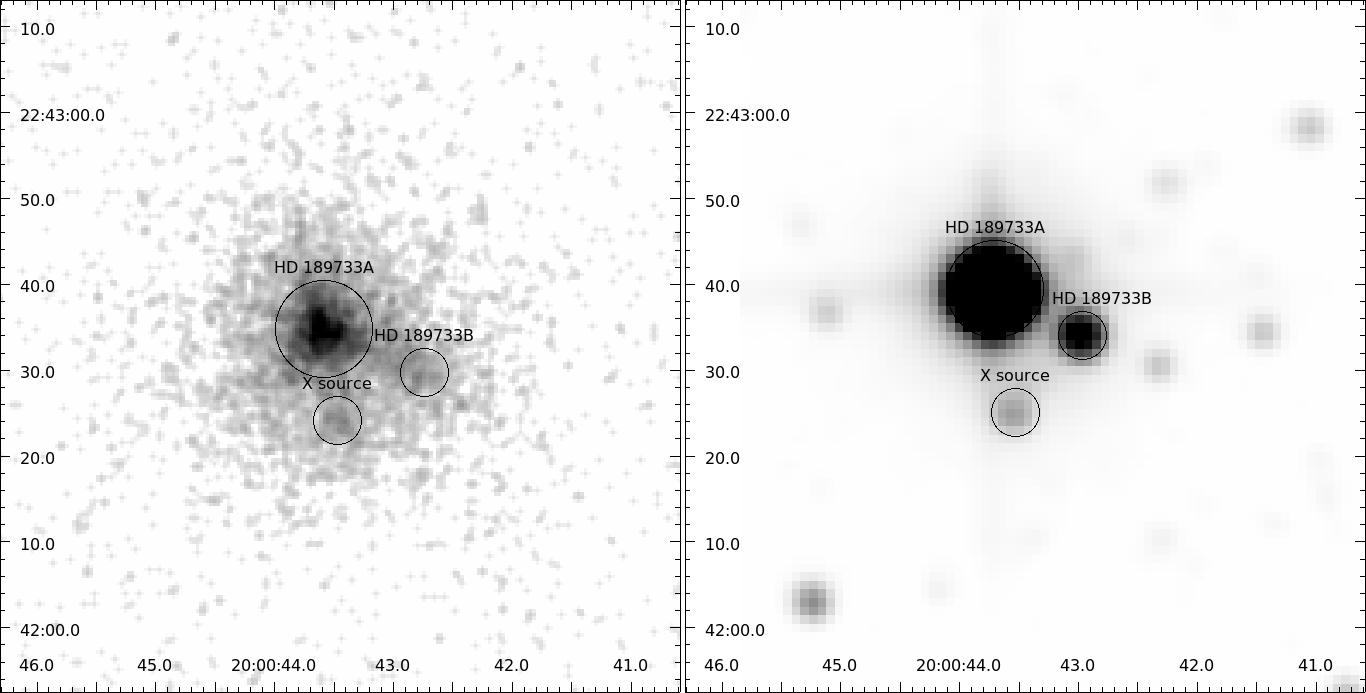}
\end{center}
\caption{\label{hd189733b} X-ray (left panel) and 2MASS-J band (right panel) images centered on \hd. 
We detect the X-ray faint companion \hd B and also the X-ray source reported in Paper I, 
both at $\sim12\arcsec$ separation from \hd~A.}
\end{figure}

In Paper II,  combining the three RGS exposures available at that time (Paper II),
we were able to perform high resolution spectroscopy in X-rays of \hd\  
and obtain density diagnostics from O~VII and Ne~IX He-like triplets.
In quiescence, the intercombination line of the OVII triplet is unusually strong, indicating
densities $> 10^{10} $ cm$^{-3}$. During the flares, the triplet is marginally consistent with a low
density. When comparing the OVII line luminosity and the ratio of the lines of OVIII/OVII species,  
\hd\ sits marginally apart from the track of Main Sequence stars. Its high density 
corona is more similar to that of active pre-Main Sequence stars.  

{\hd\ has been monitored with short {\em Swift} exposures and \citet{Lecavelier2012} 
reported a strong flare 9 hrs before a transit observed in Sept. 2011 simultaneously with 
{\em HST} and {\em Swift}.} They reported the detection of an escaping atmosphere of 
\hd b and its variability through a sequence of HST/STIS spectroscopic observations. 
They suggest that this is due to the interaction of 
the coronal activity and stellar wind of \hd A with the outer atmosphere of its planet.

\citet{Poppenhager2013} observed six primary transits of \hd b with \chandra.
They reported the detection of the planetary transit in X-rays with a 
significance level of $\ge2\sigma$ (98\% level), and inferred an inflated exoplanet atmosphere 
larger in X-rays than in the optical band. 
Despite the longer total exposure time (150ks) with respect to the time devoted to observe the 
secondary transits, 
None of the six \chandra\ exposures (for a total of 150 ks) nor the XMM exposure at the transit ($\sim50$ ks)
show bright flares comparable to those seen after secondary transits, {while some degree of time variability 
at a level of {$1-2\sigma$} is detected during these transits.}

\section{Observations and data analysis}
We obtained an \xmm\ exposure of 61.5 ks with the EPIC camera and  the {\em Medium} filter on May 7th 2012.
The observation encompasses the range of phases $0.45-0.75$, which includes a secondary transit of the planet 
and subsequent phases. The setup of the observation was chosen identical to that of our two previous
\xmm\ observations. Fig. \ref{hd189733b} shows the X-rays image obtained with the present \xmm\ observation, 
and the same sky region in the 2MASS-J band.
The {\em Observation Data Files} (ODF) have been processed with the \xmm\ {\em  Science Analysis System} 
(SAS) software ver. 12.0 to obtain event files with calibration of energy and astrometry. 
Further filters were applied to select only single and double events (PATTERN $\le 4$) and FLAG = 0 events 
as described in the SAS analysis guide.
We extracted the events related to \hd A in a circular region of $12\arcsec$ of radius,
centered on the source PSF centroid. From those events we obtained spectra and light curves. 
Arrival times of the events have been 
corrected with the SAS task {\em barycen} for timing shifts due to the satellite motion along its orbit. 

We extracted the background events in a region at the same distance from the readout node as the source 
so as to have the same detector response.
The spectra have been analyzed with XSPEC software ver 12.8 \citep{Arnaud1996,Arnaud1999}, 
using a combination of two thermal components (APEC, \citealp{Smith1999,Brickhouse2003}) 
for the best fit procedure. 
We extracted also the RGS 1 and 2 spectra of \hd A and combined them with the RGS spectra 
obtained in the previous observations to improve the count statistics and to update the results 
presented in Paper II.

Finally, we applied the wavelet analysis described by \cite{Torrence1998Practical} and previously
applied to \xmm\ light curves by \cite{Mitra2005} and \cite{Gomez2013}. 
To this purpose, we used the PN light curve binned at 50~s intervals. 
Further details on the method and the results for \hd\ are given in Sect.  \ref{wavelets}.
\begin{figure}
\begin{center}
\includegraphics[width=0.9\columnwidth]{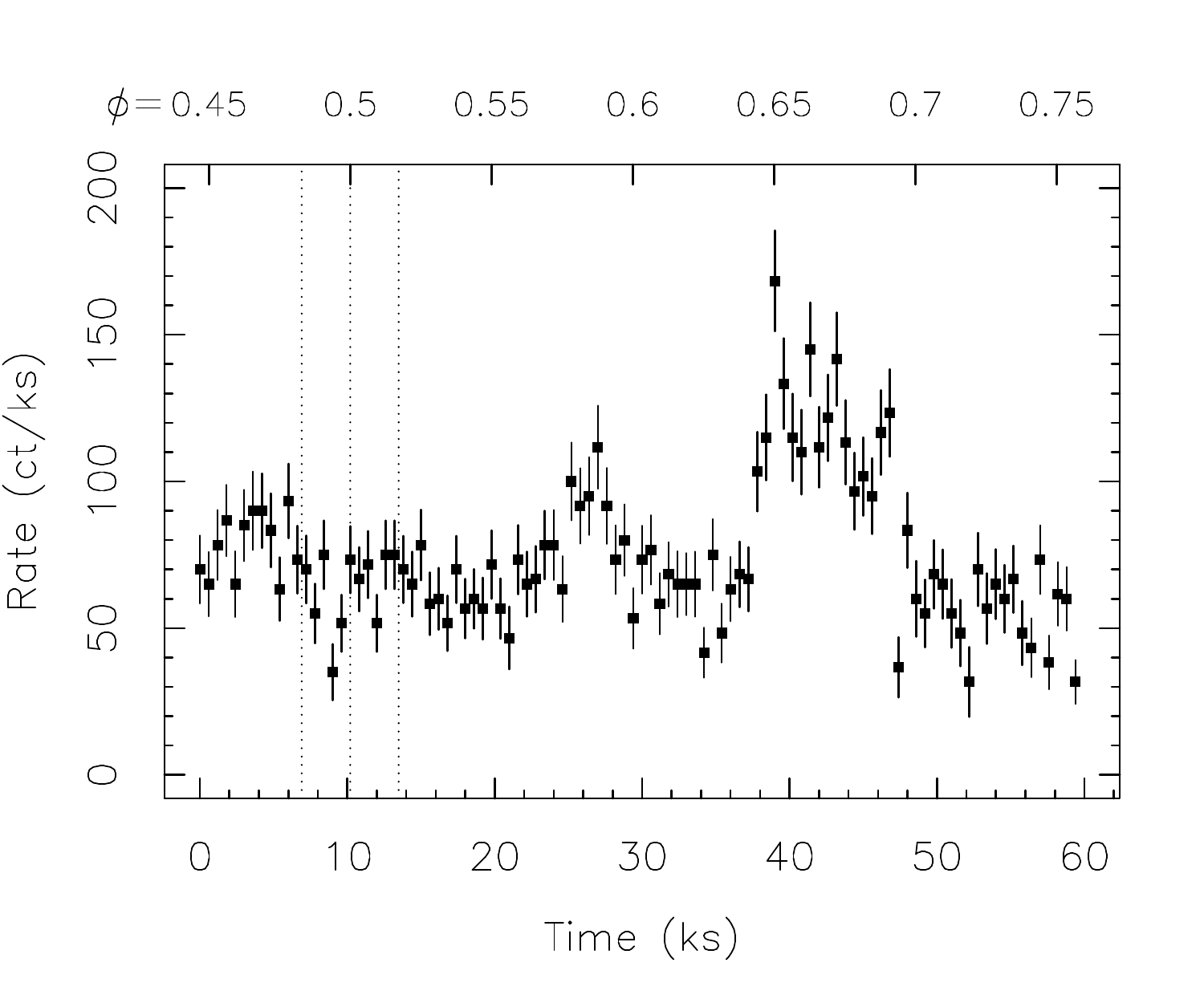}
\end{center}
\caption{\label{lcpn} EPIC PN light curve of \hd~A. Top scale reports the values of planetary orbital phases.
Vertical lines mark the ingress, middle of secondary transit and egress times. 
An intense flare is observed during the second half of the exposure time, while other minor impulses are 
apparent after the main flare and superimposed to its decay.}
\end{figure}

\begin{figure}
\begin{center}
\includegraphics[width=0.9\columnwidth]{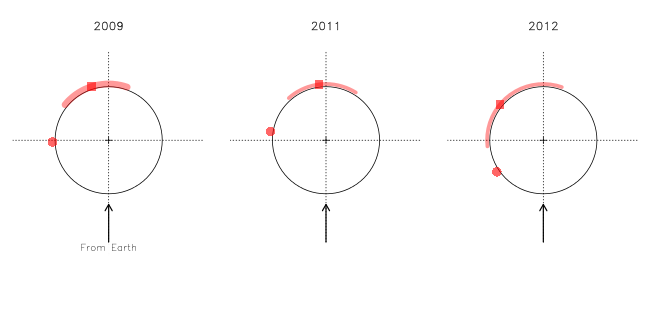}
\end{center}
\caption{\label{phases} Ranges of planetary phases observed during the \xmm\ exposures obtained in 2009,
2011, and 2012 marked on the stellar surface (polar view). We marked also the points on the surface corresponding
to the subplanetary point and the point 75$^o$ forward of it when the planet is at the phases of the three large 
flares observed in the \xmm\ exposures.}
\end{figure}

\section{Results}
\subsection{Properties of the coronal spectrum}
We first looked at the properties of the X-ray emission in terms of plasma temperatures and flux.
As we will discuss in sect. \ref{tvar}, we observed an intense flare during this observation 
(Fig. \ref{lcpn}) , 
which resembles very closely the flares recorded in the previous \xmm\ observations taken at the same planetary 
phases in 2009 and 2011.
We have performed a best-fit modeling of the spectra accumulated before the main flare ($0-36$ ks), 
during the flare ($37.5-47.5$ ks) and after the flare ($50-60$ ks). For the  
initial and final temporal intervals,
we adopted a sum of two thermal components (APEC). For the flare spectrum, 
we added a third thermal component, keeping fixed the parameters of the other two components
to the values found in the first interval, and fitting 
a scaling factor for the pre-flare model. For the interval post-flare, we tried a fit with three components,
and verified that the hottest component present during the flare was no more required for the best-fit of
the spectrum.

\begin{figure*}
\begin{center}
\includegraphics[width=0.48\columnwidth]{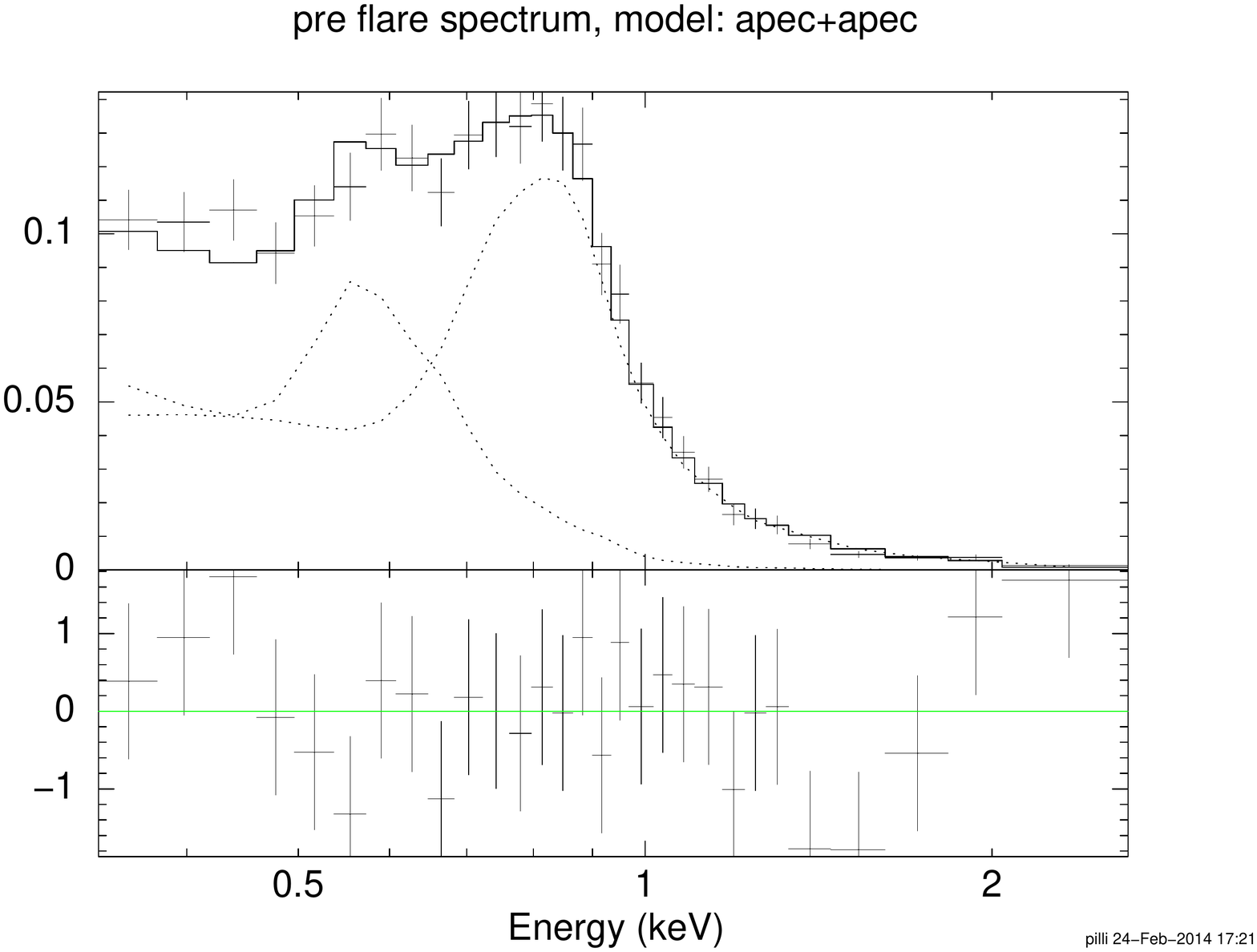}
\includegraphics[width=0.48\columnwidth]{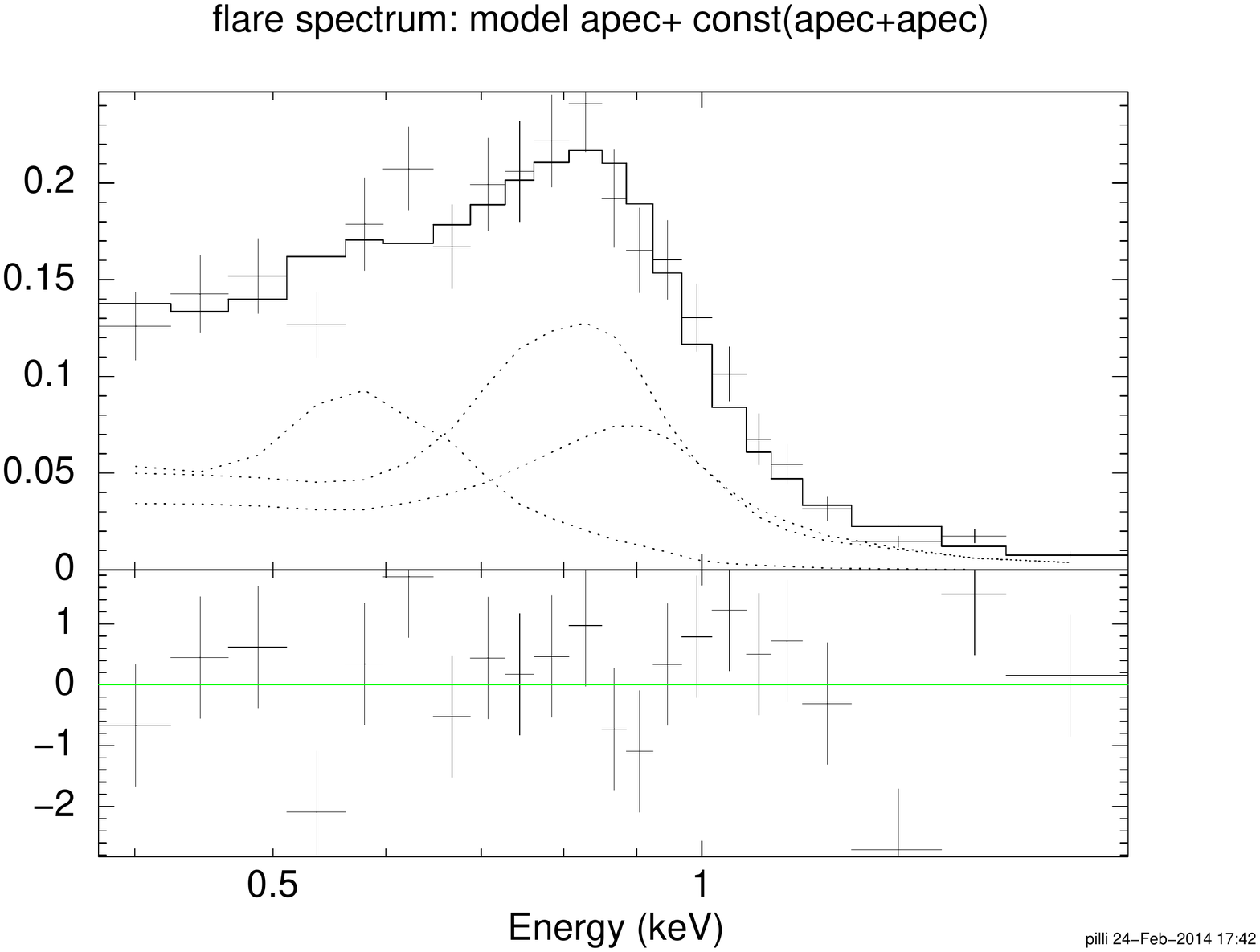}
\end{center}
\caption{\label{specpn} Spectra of PN before the flare (left panel) and during the main flare (right panel).
Dotted lines show the single thermal APEC components adopted for the best fit model.}
\end{figure*}

The results are reported in Table \ref{tabfit} and spectra with models are shown in Fig. \ref{specpn}.
the quiescent corona of \hd\  is well-represented by a thermal model with two main temperatures 
similar to the findings of  Paper II:
a cool component  with temperature $kT_1 \sim 0.2$ keV and a warm one with $kT_2 \sim 0.7$ keV, 
with a ratio of their emission measures 
$N_{cool}/ N_{warm} \sim 0.83$. During the main flare, an additional hot component ($kT_3$) 
is found, similar to the flares observed in Paper I and II. 
The flares observed in 2009 and 2011 were bright, accounting for a  doubling of $L_X$, but also cool.   
Here again the flare corresponds to an increase of the quiescent rate by a factor two, and is best represented 
by a modest ``hot" component ($kT \sim 0.8$ keV). In the post flare interval the hot component has vanished, 
and the spectrum is again well described by the cool and warm components of the pre- flare interval. 
The quiescent PN spectrum recorded in 2009 was described by two thermal components with $kT \sim 0.2$ 
keV and $kT \sim 0.5$ keV, respectively.
\begin{table*}[!t]
\caption{\label{tabfit} Parameters of best fit model to spectra of \hd\ before and during the main flares 
in 2012, 2011, and 2009. Best fit of 2007 spectrum with two temperatures model is also reported.  
{Relative errors of luminosities and fluxes are estimated to be around 10-15\%. 
We assumed a distance to the star of 19.3 pc \citep{Bouchy05}.} 
}
\resizebox{\textwidth}{!}{
\begin{tabular}{l c c | c c c c c c c c}\hline\hline
Year & Phases & Interval & kT$_1$  & kT$_2$ & kT$_3$ & E.M.$_1$  & E.M.$_2$ & E.M.$_3$ & log $f_X$ & log $L_X$  \\
&       &  (keV)  &  (keV) & (keV)  & (cm$^{-3}$)   & (cm$^{-3}$)  &  (cm$^{-3}$) & (erg s$^{-1}$ cm$^{-2}$) & (erg s$^{-1}$) \\\hline
2012 & 0.45-0.64 & Pre    & 0.19 $_{-0.03}^{+0.03}$ & 0.7 $_{-0.03}^{+0.03}$ & -- & 10.7$_{-1.3}^{+1.3}$ & 8.9$_{-1.3}^{+1.3}$ & -- &  -12.6  & 28.07 \\ 
& 0.64-0.70 & Flare & (0.19) & (0.7) & 0.82$_{-.06}^{+0.06}$ & (10.7) & (8.9) & 7.60$_{-0.7}^{+0.7}$ & -12.37 & 28.27 \\
& 0.70-0.77 & Post  & (0.19) & (0.7) & 0.72$_{0.13}^{0.13}$ & (10.7) & (3.6) & 1.3$_{-0.8}^{+0.8}$ & -12.54 & 28.11 \\\hline
2011 & 0.41-0.52 & Pre    & 0.24 $_{-0.03}^{+0.02}$ & 0.73 $_{-0.11}^{+0.08}$ & -- & 5.8 $_{-0.9}^{+1.0}$ & 3.6$_{-0.9}^{+0.9}$ & -- &  -12.5  & 28.17 \\ 
 & 0.52-0.55 & Flare & (0.24) & (0.73) & 0.9$_{-.1}^{+0.1}$ & (5.8) & (3.6) & 3.0$_{-0.5}^{+0.5}$ & -12.36 & 28.29 \\
 & 0.55-0.62 & Post  & (0.24) & (0.73) & 0.62$_{0.2}^{0.2}$ & (5.8) & (3.6) & 1.35$_{-0.03}^{+0.03}$ & -12.41 & 28.24 \\
\hline
2009 & 0.45-0.54 & Pre & 0.18$_{-0.08}^{+0.08}$ & 0.47$_{0.08}^{+0.08}$ & -- & 4.1$_{-1.8}^{+1.8}$ & 5.6$_{-2.2}^{+2.3}$ & -- & -12.50 & 28.15 \\
& 0.54-0.60& Flare    & (0.18) & (0.47) & 0.99$_{-0.08}^{+0.08}$ & (4.1) & (5.6) & 3.2$_{-0.5}^{+0.4}$ & -12.37 & 28.29 \\
& 0.60-0.64& Post     & (0.18) & (0.47) & 0.75$_{-0.17}^{+0.17}$ & (4.1) & (5.6) & 1.3$_{-0.4}^{+0.3}$ &  -12.4 & 28.22 \\\hline
2007 & 0.84-1.13 & --  & 0.24$_{-0.01}^{+0.01}$ & 0.71$_{0.03}^{+0.04}$ & -- & 4.7$_{-0.3}^{+0.4}$ & 2.8$_{-0.3}^{+0.3}$ & -- & -12.60 & 28.05 \\
\hline \hline
\end{tabular}
}
\end{table*}

\subsection{RGS data \label{rgs}}
\begin{figure}
\includegraphics[width=0.9\columnwidth]{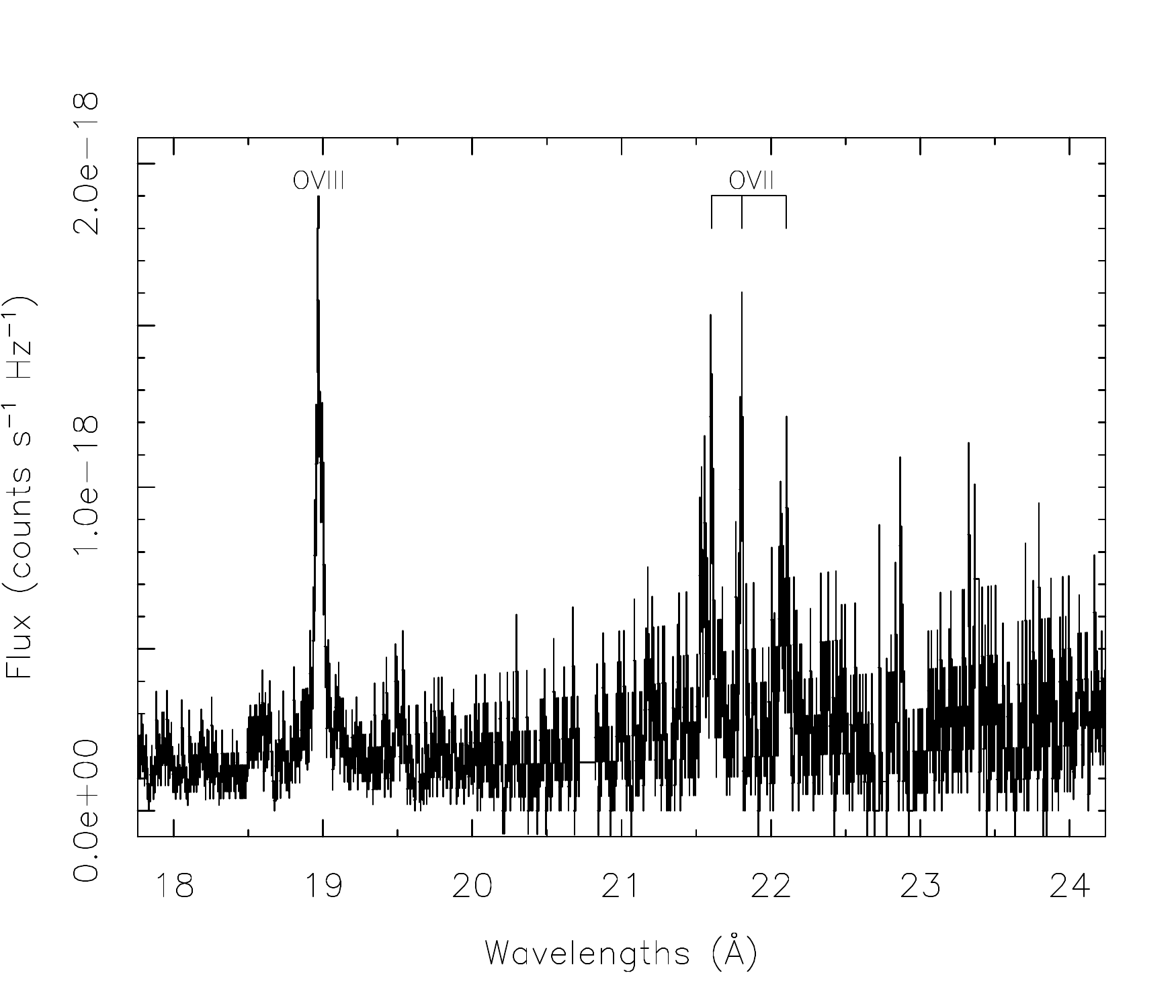}
\caption{\label{rgssplot} Portion of the sum of RGS 1 and RGS 2 spectra of all \xmm\ observations. 
The triplet of OVII and Ly$\alpha$ of OVIII are marked. 
The analysis in sect. \ref{rgs} is based on these lines.}
\end{figure}

\begin{figure}
\includegraphics[width=0.9\columnwidth]{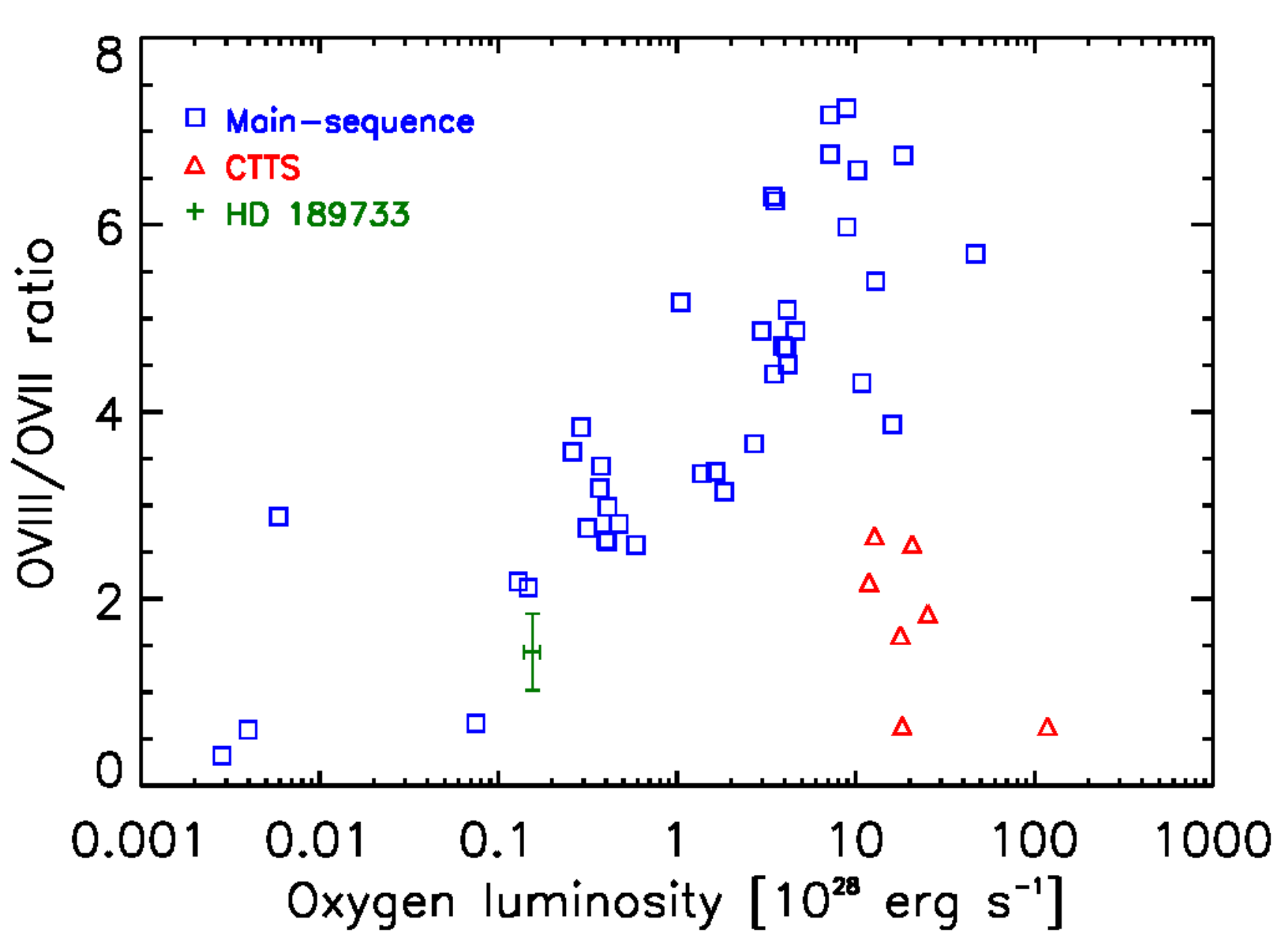}
\caption{\label{osevenplot} OVII/OVIII ratio vs. O luminosity. 
Data of Main Sequence stars and Classical T-Tauri stars (CTTS) are taken from \citet{Robrade2010} 
and \citet{Ness2004}. { The closest points above the HD189733 point are two measurements of $\epsilon$ 
Eridani (a K2 dwarf, very similar in spectral type to HD 189733), the point below and to the left belongs to Procyon, 
which is a quite inactive F5 subgiant star, and is thus more massive than HD189733.}
}
\end{figure}

The 2012 exposure allows us to add the RGS spectra to the previous datasets and refine the analysis 
already performed in Paper II. {The spectra of the different epochs show similar OVII line strengths,} 
hence we decided to sum them to increase the count statistics. 
The combined exposure time of all the spectra of RGS detectors
is $\sim180$ ks, for a total of almost 20,000 RGS counts. A portion of the spectrum around the
OVIII and OVII lines is shown in fig. \ref{rgssplot}, while the fluxes of the lines we measured
are given in table \ref{rgstable}. The method used for measuring line fluxes is the same adopted 
in \citet{Robrade2010}.
Fig. \ref{osevenplot} shows the ratio OVIII/OVII vs. O lines luminosity adapted from \citet{Robrade2010}
and from \citet{Ness2004}. 
From the intensity ratio of the OVII triplet lines $r, f, i$   
we have measured the electron density $n_e$ (\citealp{Gabriel1969}) resulting in a density at 
$1\sigma$ confidence range of $3-10\times10^{10}$ cm$^{-3}$ ($1.6-13 \times10^{10}$ cm$^{-3}$ at 
90\% confidence range).
These values are consistent with those derived in Paper II and reveal a quite dense corona, slightly
in excess respect to the coronal density values of Main Sequence less active stars.
For comparison, the electron density in the solar corona is about of $_e \sim 10^9$ cm$^{-3}$, and reaches 
values up to a few $10^{11}$ cm$^{-3}$ only during bright flares \citep{Aschwanden04, Reale2007}.   

\begin{table}
\caption{\label{rgstable}List of measured fluxes for a number of relevant lines we measured in the sum of RGS spectra.
Luminosities are in units of $10^{26}$ ergs/s (assuming a distance of 19.3 pc), photon
flux is in $10^{-6}$ photons/s/cm$^2$. Errors are given at 1$\sigma$ significance.}
\begin{center}
\begin{tabular}{lcccc}\hline \hline
Line  & Wavelength & $L_X$ & photons & photon flux \\
      & \AA\       &  $10^{26}$erg/s &   \#     & $10^{-6}\times$ph s$^{-1}$ cm$^{-2}$ \\ \hline
O VII r &         21.60 & ${2.7}^{0.5}_{-0.4}$  &  94.4 &   6.6 \\
O VII i &         21.81 & ${1.4}^{0.4}_{-0.3}$  &  45.4 &   3.3 \\
O VII f &         22.10 & ${2.0}^{0.4}_{-0.4}$  &  66.2 &   4.9 \\
O VIII Ly$\alpha$&18.97 & ${9.4}^{0.7}_{-0.7}$  & 358.3 &  20.0 \\
Ne X Ly$\alpha$ & 12.13 & ${1.2}^{0.5}_{-0.5}$  &  36.0 &   1.6 \\
Fe XVII &         16.78 & ${3.9}^{0.6}_{-0.5}$  & 153.0 &   7.4 \\
Fe XVII &         17.05 & ${3.1}^{0.8}_{-0.7}$  & 120.1 &   5.9 \\
Fe XVII &         17.10 & ${0.7}^{0.8}_{-0.8}$  &  26.7 &   1.3 \\\hline
\end{tabular}
\end{center}
\end{table}

\subsection{The detection of the stellar companion.}
The 60 ks of the 2012 \xmm\ observation represent a significant improvement of the count statistics over the 
shorter 2009 and 2011 exposures, and allow us to detect the weak X-ray emission of the stellar 
M-type companion of \hd in the 2012 images.
Fig. \ref{hd189733b} shows the X-ray image and the 2MASS J band image centered on the \hd\ system.
The detection in X-rays of \hd B has been already reported by \citet{Poppenhager2013} with an 
X-ray luminosity of $L_X = 4.7\times10^{26}$ erg/s, and is fully consistent with our upper limit 
estimate given in Paper I.
We performed a source detection process based on wavelet convolution and a multi-scale 
approach as in \citet{Pillitteri2013,Pillitteri2010aa,Pillitteri2004,Damiani2003}. 
We used the sum of the two MOS images only. This choice is motivated by  the smaller pixel 
size  of MOS cameras compared to PN camera ($2\arcsec$ vs. $5\arcsec$, respectively) that allows us
to resolve the faint companion. We detect \hd B with a rate of 1.3 ct/ks.  
We also recover the X-ray source described in Paper I south of \hd~A and characterized by a 
hard X-ray spectrum.
The X-ray count statistics of \hd B are poor given its low flux, 
thus we modeled the spectrum with a thermal emission with a component of $kT = 0.3$ keV.
Interstellar absorption is almost absent as verified with the spectral analysis of \hd A, 
due to the short distance to the system. 
With PIMMS\footnote{http://heasarc.gsfc.nasa.gov/docs/journal/pimms3.html} (ver. 4.6c) 
we obtain a flux of $7.7\times 10^{-15}$ erg s$^{-1}$ cm$^{-2}$ and a 
luminosity of only $L_X = 3.4\times10^{26}$ erg s$^{-1}$. The flux and the luminosity would 
be $9.7\times 10^{-15}$ erg s$^{-1}$ cm$^{-2}$ and  $L_X = 4.3\times10^{26}$ erg s$^{-1}$
in the case we use $kT = 0.1$ keV. 
These values are fully consistent with the upper limit given in  Paper I
and with \citet{Poppenhager2013},  confirming the low activity and the old age of this M4 type star 
and of the entire binary system. 
Based on comparison to the sample of M stars studied by \citet{Feigelson04}, 
we estimate an age of $\ge 3$ Gyr for the \hd\ system.  
\citet{Guinan2013} discussed the age of \hd\ system and placed its 
age between 4 Gyr and 8 Gyr based on H$\alpha$ emission and other activity indicators. 
The high activity of the primary star and the gyrochronology estimate of its age are thus at odds with 
the age of the M companion. This discrepancy can be solved by making the hypothesis that
the tidal interaction of planet and parent star has led to a transfer of angular momentum from the planet 
to the star, leading to rotation rate faster rotation than expected for its age,
and thus a boosted stellar magnetic activity level.  
{\citet{Lanza2010b} explained the excess of rotation of stars hosting hot Jupiters as due to
the interaction of the stellar and planetary magnetospheres that shapes a stellar corona with predominant
closed magnetic field lines, hence reducing the rate of stellar wind and the losses of
angular momentum during the Main Sequence phase.
Similar conclusions where reached by \citet{Cohen2010} based on detailed MHD numerical models.
}    

\subsection{Time variability\label{tvar}}
 The light curve of \hd\ obtained with the PN camera is shown in Fig. \ref{lcpn}. 
We have marked the phases in the top axis and the time of the planetary ingress, 
middle of secondary transit and egress with vertical lines. {In addition, Fig. \ref{phases}
shows three panels with a schematic of the star and the planetary phases observed with \xmm\
in 2009, 2011,  and 2012.}  
We note a strong variability in the phases subsequent the planet's occultation, in a striking similarity with 
what was observed in the observations taken in 2009 and 2011 (cf. Paper I and II). In particular we
notice a small flare at $\phi=0.58$,  and an intense flare at $\phi=0.65$. Both flares have a decay time
of roughly $\sim10$ ks. During the decay of the large flare, small secondary bursts  
are visible superimposed to the main decay. 
The flickering during the decay was also noticed in the 2009 flare, 
and  is reminiscent of the secondary flares occurring in large arcades on the Sun, where after the
first ignition, several bursts are triggered in adjacent loops (e.g. \citealp{Aschw2001}). 
The main flare has a peak count rate of about $1.8-2$ times above the average rate at phases just 
before the flare showing a strict similarity with the flares recorded in 2009 and 2011.


\subsubsection{Flare oscillations}
\label{wavelets} Following \citet{Zaitsev1989}, the relative amplitude of an oscillation 
of plasma density in a loop ($\frac{\Delta I}{I}$) that is triggered by the evaporating plasma after 
a flare is determined by the additional energy from filling the magnetic flux tube with hot 
plasma \citep{Mitra2005}:
\begin{equation}
\label{eq1}
\frac{\Delta I}{I} \simeq \frac{4 \pi n k_\mathrm{B} T}{B^2}
\end{equation}
where $n$ is the density of the plasma where the oscillation is occurring, $T$ is the plasma 
temperature and $B$ is the magnetic field. Thus, by measuring the relative amplitude of an 
oscillation one may have an independent measure of the strength of the magnetic field in the loop.

This scenario was assumed by \citet{Mitra2005} for a flare observed 
in the M dwarf AT~Mic. During the flare, AT~Mic experienced a magneto-acoustic oscillation
that was detected in its {\em XMM-Newton} light curve, showing $\Delta I / I \sim 0.15$. 
With a temperature $T \simeq 24$~MK and a density $n \simeq 4 \times 10^{10}$~cm$^{-3}$, 
\citet{Mitra2005} measured a magnetic field $B = 105$~G for the 
flaring loop, which is in agreement with the results obtained from light curve modeling
(e.g. \citealp{Reale2004,Reale2007}; \citealp{LopezSantiago2010}). 

\begin{figure*}[!t]
\resizebox{\textwidth}{!}{
\includegraphics[width=0.9\columnwidth]{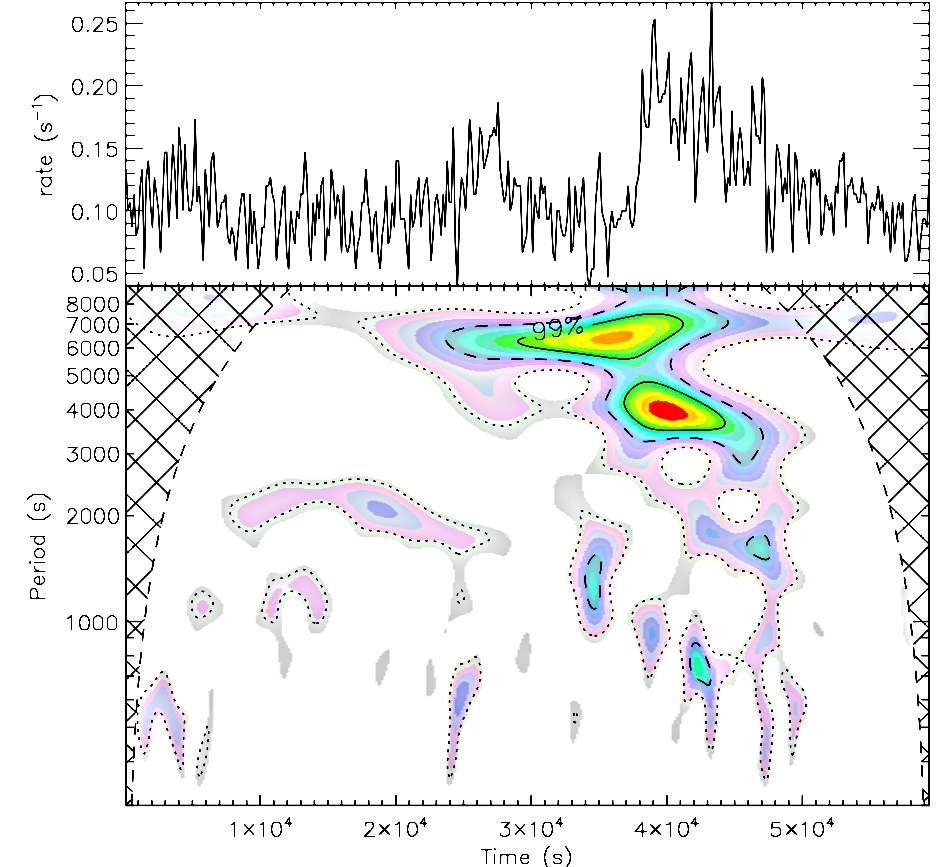}
\includegraphics[width=0.9\columnwidth]{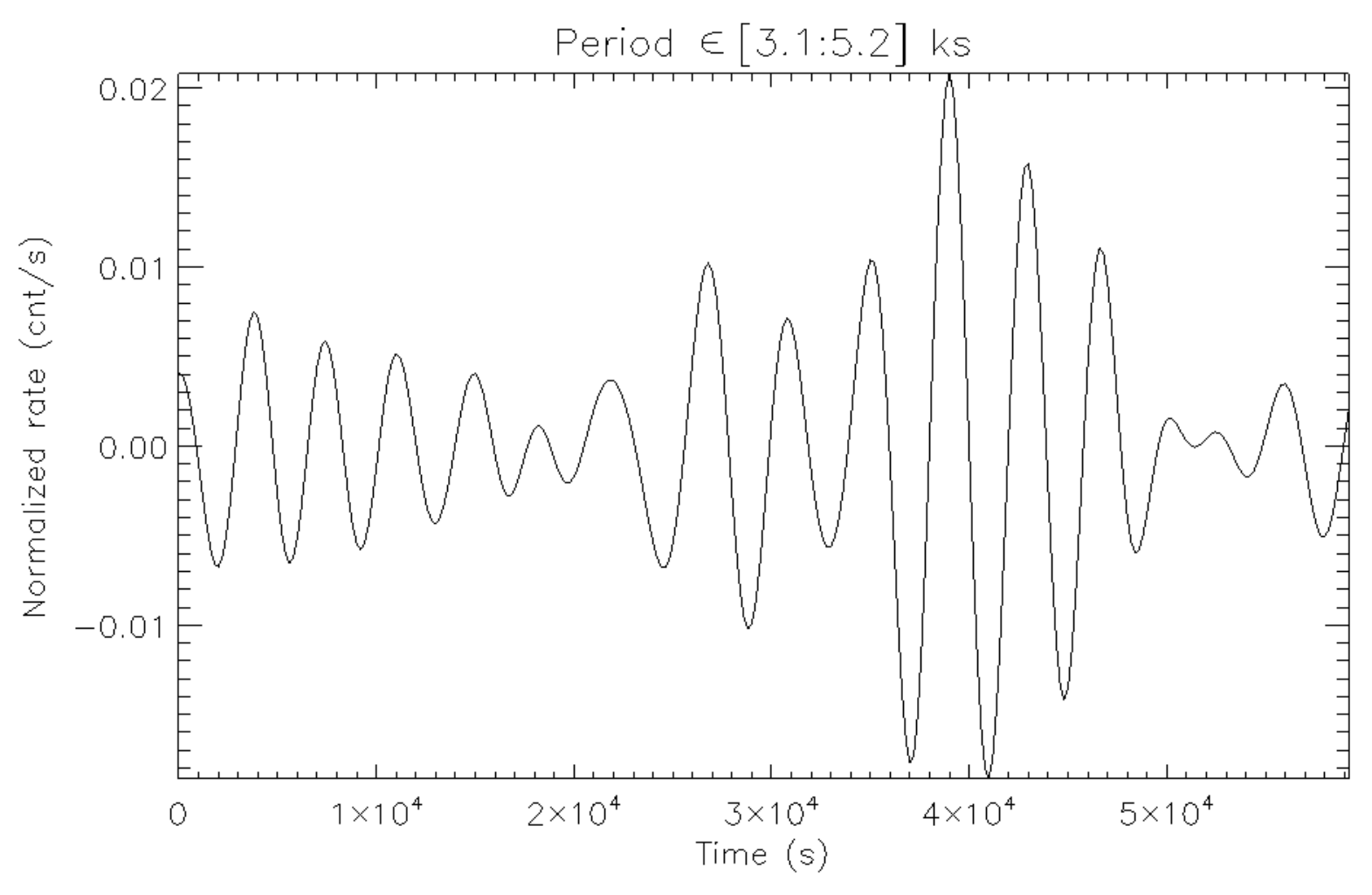}
}
\caption{Left: light curve (binned at 50 s., top panel) and period vs. time diagram (bottom panel) 
for the PN data. 
Two periods of oscillation are detected at $P \simeq 4$ ks and $P\simeq 9$ ks.\label{power}
Right panel: reconstructed light curve with the periods in $3.1-5.2$ ks.
}
\end{figure*}

{The secondary peaks after the main flare of HD 189733 resemble those observed in AT~Mic.} 
In that star, immediately after the flare peak, the light curve remains at a fairly constant 
count-rate during for approximately 8~ks and then decreases rapidly. During the lapse
in which the count-rate remains relatively constant, it shows signs of secondary peaks with periodicity 
of $\sim 3.0-4.5$~ks. 
We applied the same methodology as in \citet{Gomez2013}
to determine the significance of the detection of this oscillation. The whole PN light curve of \hd~A 
was convolved with a wavelet function that is localized in both time and frequency domains
(for our case we chose a Morlet function; \citealp{Torrence1998Practical}). 
The result of the convolution is the 2D power spectrum shown in Fig.~\ref{power} (left panel). 
The oscillation period is clearly detected at $P = 4 \pm 1$~ks 
at the 99\% confidence level, with a second (less strong) peak at $P \sim 9$~ks. The whole
light curve reconstructed with periods in the range [3.1-5.2]~ks is also shown in 
Fig.~\ref{power} (right panel). Qualitatively, this curve recovers most of the features observed 
in the real data. The figure shows a modulated signal with $\Delta I / I \sim 0.20$ during the main
flare. We assessed the significance of the peak by means of Monte Carlo simulations of constant count rate 
(equal to the mean of the observed count rate) with injected red noise as a background spectrum.
To investigate the possibility of spurious detection of high significance
peaks in the frequency domain not related with any physical process, we
created a model of light curve with a flare similar to the observed one, and injected random red
noise. Then, we applied the wavelet analysis to the resultant (noisy)
light curve. This process was repeated up to a thousand times. We did
not obtained any significant peak (with significance $> 90$\%) in any simulation
during this process other than the peak detected in the real data. 

The flare observed in 2009 has characteristics very similar to the flare of 2012. 
In that observation, but a problem with the telemetry in the data created a gap in the 
light curve during the flare's decay.
The wavelet analysis on the  2009 flare shows that an oscillation with period of $\sim5$ ks could
be present also in that flare, but the telemetry loss hampers the significance of this result.

Assuming that the source of oscillation during the flare is a magneto-acoustic wave triggered by the 
flare and using Eq.~\ref{eq1} with $n = 3-10 \times 10^{10}$~cm$^{-3}$ 
(Paper II and Sect. \ref{rgs}) and $T_{flare} \simeq 11$~MK as resulting from the analysis of the X-ray data, 
we determine a magnetic field $B = 50-100$~G. With the 90\% confidence range of $n_e = 1.6-13\times10^{10}$ 
cm$^{-3}$, we obtain $B = 40-110$~G, similar to the values of global magnetic field inferred by \citet{Fares2010}
through spectropolarimetry.

Under coronal conditions, i.e. low plasma density and strong magnetic confinement, the period of
the oscillation is proportional to the loop length ($\tau = L / c_\mathrm{s}$) for first order
mode \citep{Gomez2013}. The sound velocity in the plasma is  
$c_\mathrm{s} = 2 \gamma k_\mathrm{B} T / m_\mathrm{p}$. For an adiabatic gas 
($\gamma = 5/3$), $c_\mathrm{s} = 1.66 \times 10^4 \sqrt T$ and, hence, 
$c_\mathrm{s} = 5.5 \times 10^7$~cm\,s$^{-1}$. According to this result, the loop length 
is $L \simeq 4\times10^{11}$~cm or approximately $4 R_\star$, assuming 
$R_\star = 0.8 R_\odot$. This gives a height for the loop of $\sim  0.007$~AU. 
Assuming two foot points on the star, {while it is four times smaller than the star-planet 
separation}, this length still constitutes a significant fraction of the distance 
between the two bodies ($\sim25\%$).

Another derivation of the loop length is given by following \cite{Mathioudakis2006}. 
The period of a standing, slow-mode  oscillation is given by
\begin{equation}
\label{eq2}
P = \frac{L}{7.6 \times 10^{-2} N \sqrt T}
\end{equation}
with $P$ in seconds, $L$ in Mm and $T$ in MK, N the order of the harmonics, with N=1 being the first harmonic. 
This relation is independent of the density of the plasma and it assumes that the longitudinal speed of the tube 
is similar to  the sound velocity of the plasma, i.e. the magnetic field is dominant 
over the thermal pressure of the plasma. 
For $P = 4000$~s, $T \simeq 11$~MK and the first harmonic, $L \sim 10^{11}$~cm or
approximately $1.8 R_\star$, and thus shorter than the length determined previously, but still 
a significant fraction of the separation between the star and the planet.

In both cases a long loop, larger than the stellar radius, is involved. 
The size of the loop is larger than those 
commonly observed on the Sun, and resembles the estimated sizes of loops derived for active
stars like Proxima Centauri \citep{Reale2007} or in young active stars with disks showing powerful 
and long-lasting flares \citep{McCleary2011,Caramazza2007}). 
The origin of this loop could be related to the interplay between the magnetic field
of the planet and the magnetosphere of the star.
This would result in a reconnection event triggered in a loop elongated toward the planet.
Detailed MHD simulations by \citet{Cohen2011} show that a plasmoid can be ejected from the star
toward the planet under a favorable configuration of the global magnetic field, reinforcing the scenario of a flare event occurring in an elongated loop and triggered by the planet. 

\subsection{Variability of the X-ray luminosity}
To date, we have observed HD189733 for a total of about 130 ks between planetary phases of 0.45 and 0.7.  
We have identified three bright flares with a peak count rate of more than twice the average level before 
the flares.
Other variability at lower levels is detected in the other observations. We want to assess how this 
variability  compares with the X-ray variability of young active stars, like those of 
the clusters of Pleiades and NGC 2516, and field stars that are supposed to be coeval with \hd.
For this purpose, we have considered the PN light curves and the count rates measured
in 2007 (at the planetary transit), 2009, 2011, and 2012 observations (at the planetary eclipses). 
We calculated the cumulative distribution of the variability amplitude
of the luminosity, meaning the ratio between the luminosity in each bin of the light curve and its minimum. 
This distribution can be interpreted as the fraction of time 
during which the star, on average, is observed with a rate greater than a certain value. 
A few bins with very low count rate ($<50 ct/ks$) were discarded because they could have some problems 
related to the gaps of time intervals (GTIs) not properly corrected by SAS. 

\begin{figure}
\begin{center}
\includegraphics[width=0.99\columnwidth]{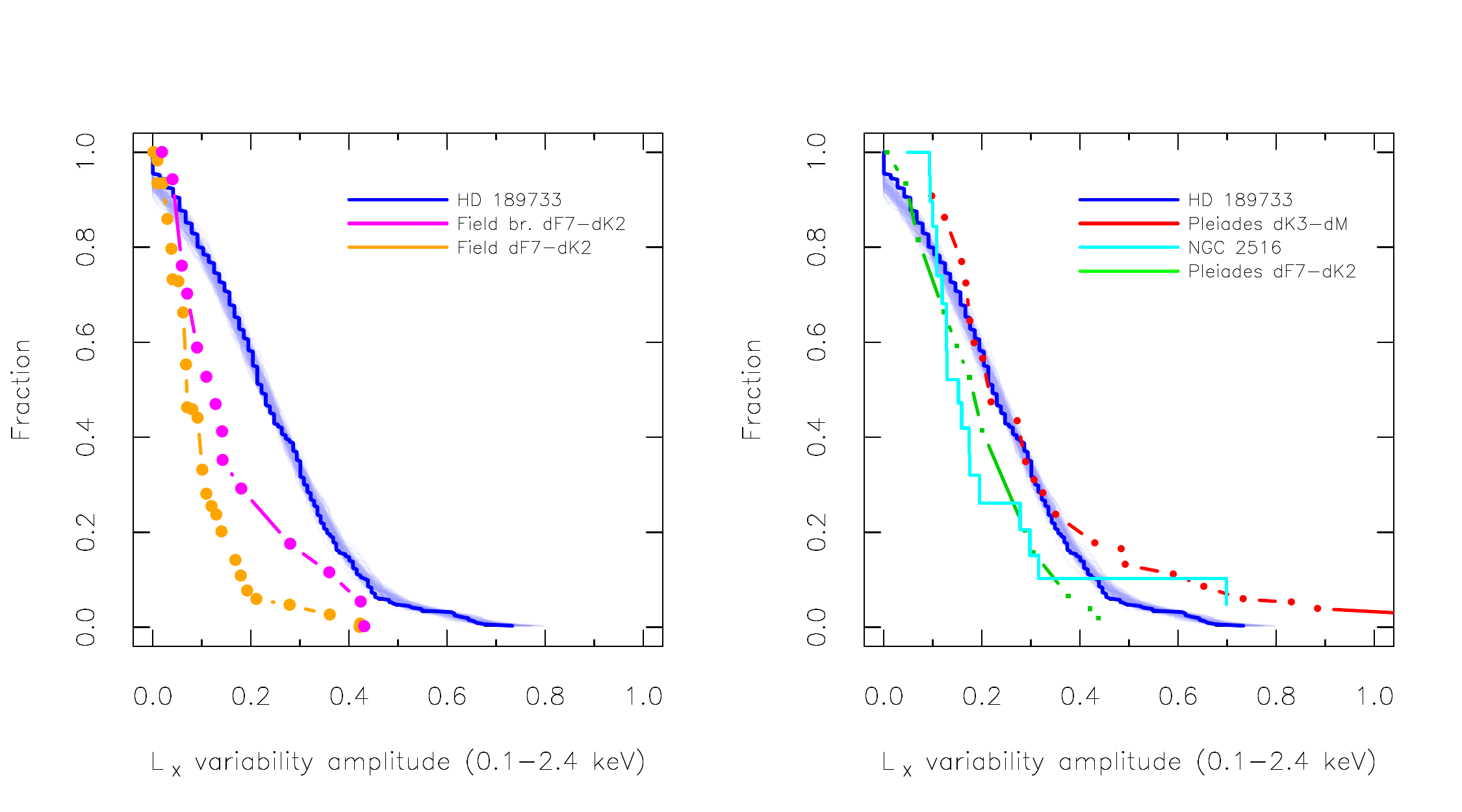}
\end{center}
\caption{\label{lxampv} Cumulative distributions of X-ray luminosity variability amplitude observed 
in \hd\ and for comparison in field stars (left panel) and young clusters  (right panel) like Pleiades 
and NGC 2516 taken from \citet{Marino2003} and \citet{Marino06}. 
Shaded areas are the Monte Carlo simulated distributions. 
For field stars we used the whole sample and the bright sample with luminosities in the range of Pleaides stars.
For NGC 2516 we use the dF7-dK2 sample of \citet{Marino06}.}
\end{figure}

In order to evaluate the effect of uncertainties in each time bin on the overall time variability 
distribution, we have adopted a Monte-Carlo procedure. We have simulated 1000 values of count rates
for each bin, with the mean and sigma equal to the observed mean and standard deviation.
From each of the 1000 realizations, we have computed a cummulative variability distribution similar to that 
of real data, and have determined the extent of the region that can be populated by those distributions 
(see Fig. \ref{lxampv}). 

At this point we have evaluated the baseline luminosity and the values of variability amplitude, 
the final step was to create a cumulative distribution of the values of variability amplitude.
We have converted the count rate distributions to fluxes and luminosities. 
Based on our previous \xmm\ observations,  we can suppose that the quiescent corona has a temperature 
average of between the two main components (0.2 keV and 0.7 keV, see Table \ref{tabfit}) 
around $kT = 0.4$ keV and during the largest flares the temperature rises up to $\sim1$ keV.
By means of PIMMS (v.4.6a), we set a 1-T APEC thermal model with $kT= 0.4$ keV and obtained a rate to flux
conversion factor (CF) that we used for the count rates $r \le 0.12$ ct/s. 
Absorption was set to zero given the short distance to the star (19.3 pc).
We made the assumption that count rates above 0.12 ct/s are due to some kind of flare activity and the plasma is hotter
than in the quiescent phases. We calculated the CF corresponding to $kT = 1$ keV, and
linearly interpolated the values of CFs onto the values of rates between 0.12 and the maximum rate.
The difference in CF between 0.4 keV and 1.0 keV is about 25\%. 
From fluxes we derived X-ray luminosities and we calculated the
cumulative distribution of variability amplitude shown in Fig. \ref{lxampv}. The distribution
of variability amplitudes tells that, on average, for about 20\% of the time the star shows 
a luminosity 1.5 times above its quiescent/baseline luminosity because of time variability.

We compare the distribution of variability amplitude with analogous curves derived by \citet{Marino06} for
NGC~2516 (\xmm\ data), and \citet{Marino2003} for the Pleiades and field stars (ROSAT-PSPC data).  
Since the latter comparison is with ROSAT data, we derived the fluxes in the ROSAT band (0.1-2.4 keV). 
Compared to field stars (Fig. \ref{lxampv}, left panel), \hd\ shows a higher variability amplitude.
In X-rays, \hd\ appears more variable than the X-ray brightest G and K type stars in the solar neighborhood. 
Field stars and dF7-dK2 stars in NGC 2516 (right panel of Fig. \ref{lxampv}) vary by up to a factor 
about 40\%, whereas \hd\ shows variability as  high as a factor 80\%. Compared with young
clusters, \hd\ is still more variable than late type stars of NGC~2516 and it looks similar to late dK stars 
of Pleiades. 
Comparing with active dM stars, we observe that these objects exhibit a higher degree of variability than \hd, 
up to a factor 20, in the 0.1-2.4 keV band \citep{Marino2000}. 
In summary, \hd\ shows a X-ray variability different to and more enhanced than in field stars, and it
resembles young active stars. 

\begin{figure}
\begin{center}
\includegraphics[width=\columnwidth]{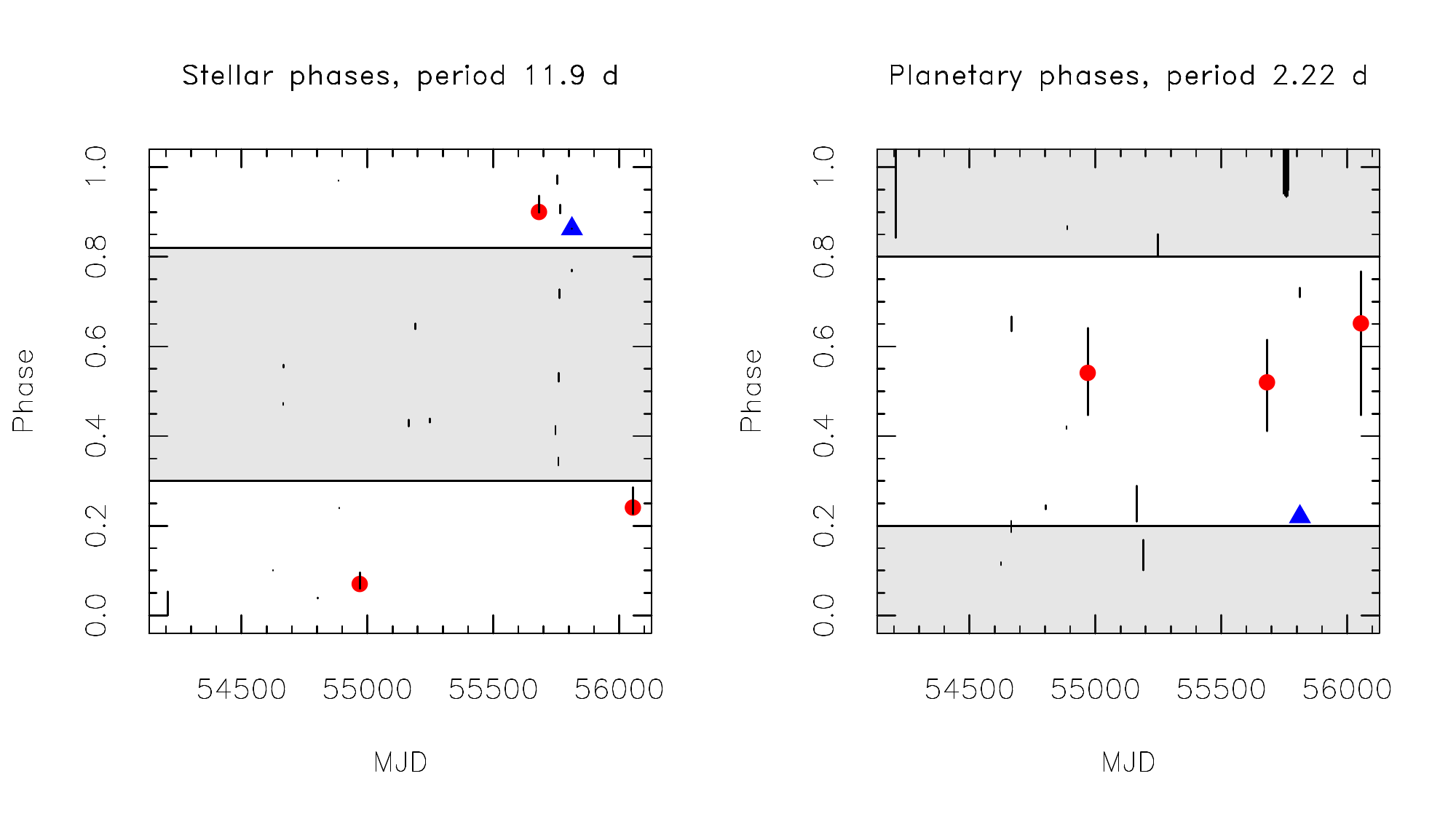}
\end{center}
\caption{\label{xphases} Stellar rotational phases (left panel) and planetary orbital phases (right panel) 
during the X-ray observations obtained with \xmm, \chandra\ and {\em Swift}. 
For the stellar rotational phases the zero point is set to the beginning of \xmm\ exposure in 2007. 
Observations with intense flares are marked with filled symbols: red solid circles for \xmm\ flares and
blue triangle for the {\em Swift} flare. In both panels, shaded areas mark the phase intervals without 
observations of large flares. {We used the {\em Swift} observations nr. 
00036406010, 00036406011, 00036406012, 00036406013, 00036406014, 00036406015, 00036406016, 00036406017, and
\chandra\ observations nr. 12340, 12341, 12342, 12343, 12344, 12345.}}
\end{figure}

\begin{figure}
\includegraphics[width=0.9\columnwidth]{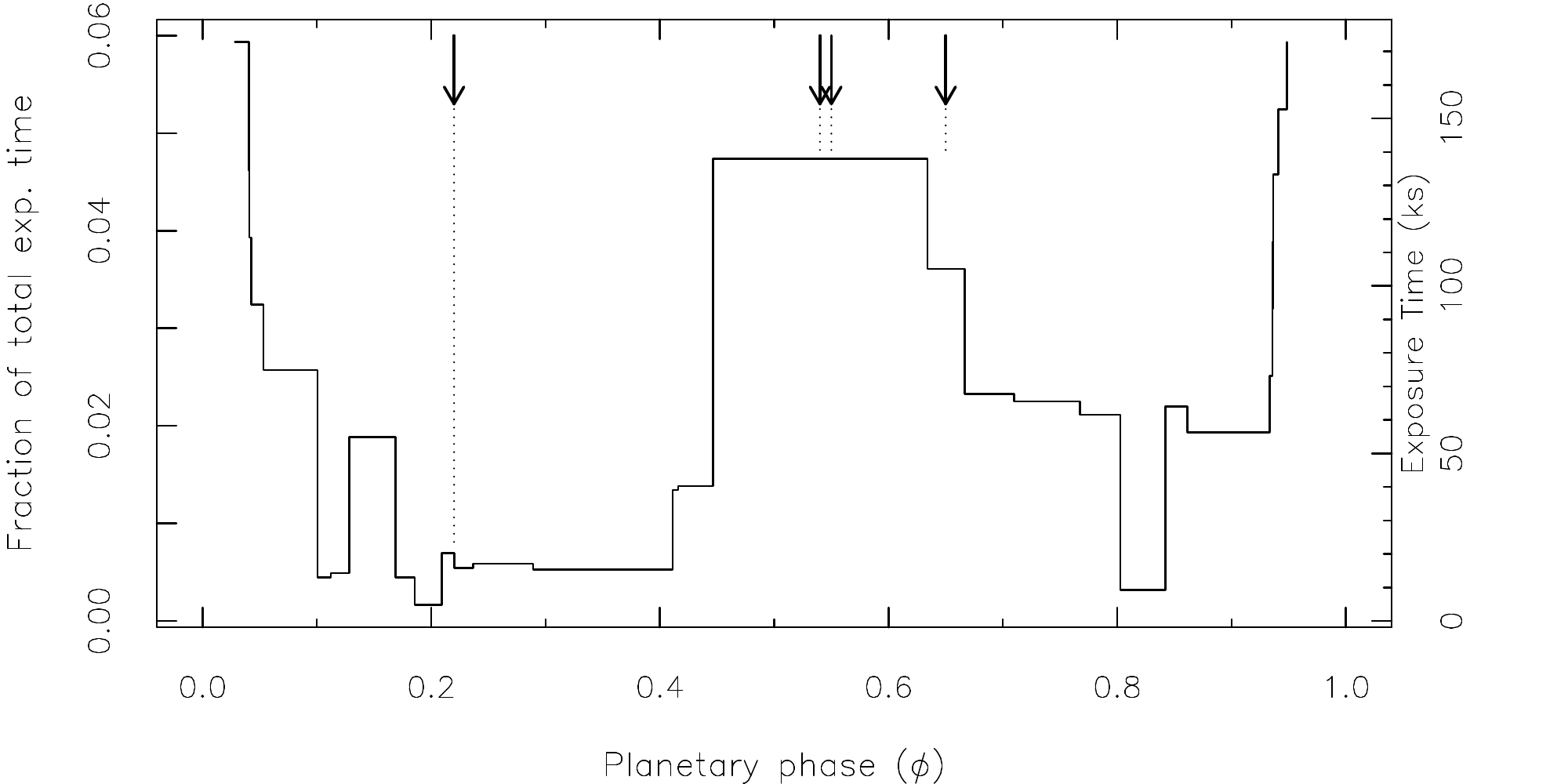}
\caption{\label{exptphi} Fractional and absolute exposure time of \xmm, \chandra\ and {\em Swift} 
devoted to observe \hd\ as function of the planetary orbital phases.}
\end{figure}

\section{Discussion}
The most interesting result of our analysis comes from the wavelet reconstruction
of the 2012 flare that leads to an estimate of the local magnetic field consistent with \cite{Fares2010},
and a long loop as the flare site. A similar finding was obtained for the 2009 flare, although at a lower 
significance. 
We can speculate that the presence of the planet can
generate a  force of tidal and magnetic nature. The passage of the close planet exerts a pulling
force on the magnetic loops and on the plasma and can create magnetic reconnection and flares on extended
loops. 
{\citet{Fares2010} have estimated the poloidal and toroidal components of the stellar magnetic field
through spectropolarimetry at three epochs (2006, 2007, and 2008). At the stellar surface, 
the magnetic field has predominantly a toroidal component of 20-40~G and some changes between
the three epochs are detected. More interestingly, given that the magnetic field at the planet orbit 
is not uniform, at orbital phases where the field is stronger is more likely for the planet to trigger
reconnection events \citep{Fares2013, Cohen2011}.
The strength of the loop magnetic field that we infer is consistent with the Fares et al. estimates. 
This makes plausible a scenario in which the passage of the planet can trigger reconnection events and 
stretching of the stellar magnetic field lines.}

The number of large flares observed in 2009, 2011, and 2012 at phases after 
the secondary transit appears to be surprising. These flares have a flux about twice the quiescent 
flux and significance above 3$\sigma$. No similar flares have been observed during the six exposures 
centered on transits, when variability at a lower level of significance and amplitude is seen 
\citep[e.g.][]{Poppenhager2013}. 
In Fig. \ref{xphases} we plot the phases of the stellar rotation and the 
orbit of the planet as function of the times of the observations obtained with \xmm, \chandra\ and 
{\em Swift}. 
The zero point of stellar rotational phases is arbitrarily set at the beginning of the 2007 \xmm\ observation.
{For this rotational period we have chosen the value of 11.9 days, 
reported by \citet{Fares2010} for the equatorial
rotational velocity. In the same paper differential rotation is reported with the high latitudes of the star
rotating at about 16.5 days. Since the stellar magnetic field has a predominant toroidal component, we
expect that the active regions populate a belt around the stellar equator and rotate with the equatorial
11.9 day period.}
The filled symbols represents the observations with strong flares detected with \xmm\ and {\em Swift} observations 
\citep{Lecavelier2012}.
The large flares are observed in a limited 
range of stellar phases, namely at $\phi = 0.85-0.2$. At the same time, the planetary phases
are at 0.55-0.65 for the three \xmm\ flares and at 0.2 for the {\em Swift} flare. 
As pointed out before, during the six \chandra\ observations and the other ten {\em Swift} snapshots, 
such large flares are not seen, so the flares seem to be seen only during half of the planetary phases
and half of the stellar rotational phases which driver remains unclear. 
The stellar rotational phases at which the large flares occur map into stellar longitudes, 
so that they correspond to the range $\simeq300-70\deg$.

In Fig. \ref{exptphi} we show the fractional and the absolute exposure time 
of \xmm, \chandra\ and {\em Swift} as a function of the planetary phases. 
The sum of all the exposure times is $\sim369.6$~ks.
The arrows mark the phases where the strong flares occurred. 
Most of the time has been devoted to observing \hd\ at the planetary transits. Under the hypothesis
that the strong flares are uniformly distributed with respect to the planetary phases, it is expected 
to observe more flares where the exposure time is higher, so that the fractional time vs. phases can 
be read as a density probability function of observing the flares. 
Table \ref{expectedflares} gives the expected number of flares 
{based on the density probability distribution
obtained from the exposure times vs. $\phi$ curve}, 
the observed number of flares and their significance level in four different phase ranges.
We observe both an excess of flares at $0.45-0.65$ phase interval plus a deficit in the $0.9-0.1$ range 
with significance of $>1\sigma$, while the {\em Swift} flare is within $1\sigma$ of the expected 
number of flares.
 
{Here we find a suggestive analogy with the short term variability 
observed in Ca II lines by \citet{Shkolnik08}  around phases $\phi=0.7-0.8$.  When the planet
is at phase 0.55 a region on the surface leading about $70-75\deg$ the subplanetary
point emerges from the limb. This region could be the site of the X-ray flares
and responsible for the enhanced chromospheric activity observed in Ca II lines. 
Given the motion of the planet, it is expected that the active region should 
travel at the same rate of the orbital velocity of the planet on the stellar surface, 
and produce strong flares while on the visible face of the star, 
or about when the planet is in the range $\phi=0.65-0.15$.
However, X-ray variability as seen at $\phi=0.55-0.65$ is not observed around $\phi=0$.}

Although there in no conclusive evidence of a significant excess of flares after the secondary transits, 
we still suggest that the planet might trigger such flares only  when it passes close 
to regions of locally high magnetic field on the underlying star at particular 
combinations of stellar rotational 
phases and orbital planetary phases.

\begin{table}
\caption{\label{expectedflares} Number of expected and observed flares, and their significance in four
planetary phase ranges.} 
\begin{tabular}{l r r r }\\
\hline \hline
Phase range  & Expected & Observed & Significance \\\hline
0.45 - 0.65 & 1.6      & 3        &  1.1  \\
0.9  - 0.1  & 1.6      & 0        &  1.3  \\
0.1  -0.45  & 0.4      & 1        &  1    \\
0.65 - 0.9  & 0.4      & 0        &  0.6  \\
\hline
\end{tabular}
\end{table}

The hypothesis of a locally enhanced magnetic field within a range of stellar longitudes agrees with maps of 
the stellar magnetic field \citep[Fig. 12 in ][]{Fares2010} and the estimate of the magnetic
field in the flaring loop obtained by our wavelet analysis in Sect. \ref{wavelets}. 
In particular, the configuration of the magnetic field in 2008 given by \citet{Fares2010} shows also 
two ranges of maxima of the magnetic field with respect to the stellar longitudes. It is also noteworthy 
that the configuration has changed from 2007 to 2008, opening to the possibility that \hd\ could have 
activity cycles on a time scale of a few years. 

Our X-ray observations suggest that the configuration of the magnetic field of \hd\ has  
a range of stellar longitudes with enhanced field, and that the passage of the planet  over these
longitudes could trigger high flaring activity. 
The wavelet analysis of the PN light curve predicts the presence of an extended loop  
a few stellar radii in length, as derived in the case of active young stars and reinforces the idea that 
the close-in planet could play a role in triggering extra activity in such structures.  
The intensity of the magnetic field is of the same order of the global dipolar field of the star.
This fact supports the idea that the geometry of the flaring structure is thin and elongated  
rather than be a small scale structure wrapped around very intense magnetic field close the stellar
surface. 
The way to prove or confute this scenario relies on observations of other intense
flares where the wavelet analysis could be applied. 

The variability amplitude of \hd\ is higher than in field stars in the ROSAT band ($0.1-2.4$ keV), 
at odds with an old age of the star and more similar to young G-K type Pleiades. 
On the other hand, the detection of the M4 stellar
companion at only $4.3\times10^{26}$ erg/s puts a lower limit on the age of the system at $t>3$ Gyr. 
To date, \hd\ has been observed in X-rays for 170 ks with \xmm, 150 ks with \chandra\ and 50 ks with {\em Swift}.
The presence of four intense flares with releases of energy of $\sim10^{32}$ ergs 
observed in a total exposure time of $\sim370$ ks is remarkable. 
Very young stars in Orion Nebula Cloud have a flare frequency of 
only one large flare every $\sim500$~ks \citep{Wolk2005,Caramazza2007}, although those flares have a higher 
energy budget of about $10^{34}-10^{35}$ erg. 

In summary, the excess of flaring activity, the size of the loop involved, and the X-ray variability amplitude 
in \hd\ could be interpreted as consequences of star-planet interaction of magnetic origin. 
The presence of the planet would induce extra time variability and coronal activity that the star alone 
could not manifest. 

\section{Conclusions}
In this paper we have reported on the third \xmm\ observation of \hd\ at the eclipse of the planet.
The main findings of this study are:
\begin{itemize} 
\item We observed a third strong flare after the secondary transit of the planet. 
\item A wavelet analysis of the light curve reveals that in this case the flaring structure 
may be as big as four stellar radii. The magnetic field in this loop is in the range 40-110 G, 
in agreement with the estimates of global magnetic field of the star 
derived from spectropolarimetry \citep[see][]{Fares2010}. 
\item This large length suggests an origin due to magnetic interaction between the star and the close in planet.
Qualitatively speaking, a magnetic field associated with the planet can exert a force 
on the plasma and the coronal loop when the planet passes close to regions of the stellar surface 
with enhanced magnetic field.
\item The detection of the M type companion \hd B at level of $3.4\times10^{26}$ erg/s confirms the very old 
age of this star, at odds with the age of the primary estimated from gyrochronology.
\item The discrepancy of age between \hd A and \hd B hints a tidal interaction of the main star with its 
hot Jupiter and the transfer of angular momentum.
\item The updated analysis of RGS data confirms the dense corona of \hd\ similar to more X-ray luminous 
and active stars. 
\end{itemize}

In the introduction we have discussed that star-planet interaction could be either of gravitational/tidal or 
magnetic origin. The difference of age estimates from magnetic activity indicators between the primary and secondary 
components of the \hd\ system 
implies that a tidal transfer of angular momentum must be occurring.

The additional suggestion here is that big flares are seen at specific planetary phases and stellar longitudes. 
More importantly, the discovery of a magnetic loop of the order of a few stellar radii and thus a significant
fraction of the star-planet separation, provides a very strong, although not decisive, evidence of magnetic
star planet interaction as well. 
\acknowledgments{IP acknowledges financial support of the European Union
under the project ``Astronomy Fellowships in Italy" (AstroFit).
S.J.W. was supported by NASA contract NAS8-03060.
}

{\it Facilities:} \facility{\xmm}.

%
 \end{document}